\documentclass[english,aps,manuscript]{revtex4-1}
\usepackage{charter}
\usepackage[T1]{fontenc}
\usepackage[latin9]{inputenc}
\usepackage[letterpaper]{geometry}
\usepackage{color}
\usepackage{amsmath}
\usepackage{graphicx}
\usepackage{esint}

\makeatletter
 
 \@ifundefined{textcolor}{}
 {%
   \definecolor{BLACK}{gray}{0}
   \definecolor{WHITE}{gray}{1}
   \definecolor{RED}{rgb}{1,0,0}
   \definecolor{GREEN}{rgb}{0,1,0}
   \definecolor{BLUE}{rgb}{0,0,1}
   \definecolor{CYAN}{cmyk}{1,0,0,0}
   \definecolor{MAGENTA}{cmyk}{0,1,0,0}
   \definecolor{YELLOW}{cmyk}{0,0,1,0}
 }

\makeatother

\usepackage{babel}
\begin{document}

\title{Crater functions for compound materials: a route to parameter estimation
in coupled-PDE models of ion bombardment}

\author{Scott A. Norris}

\address{Southern Methodist University\\
Department of Mathematics\\
3200 Dyer Street \\
Dallas, TX, 75275}

\author{Juha Samela}

\author{Matias Vestberg}

\author{Kai Nordlund}

\address{University of Helsinki\\
Department of Physics\\
PB 43, 00014 University of Helsinki\\
Finland}

\author{Michael J. Aziz}

\address{Harvard University\\
School of Engineering and Applied Sciences\\
29 Oxford Street\\
Boston, MA 02138}
\begin{abstract}
During the ion bombardment of targets containing multiple component
species, highly-ordered arrays of nanostructures are sometimes observed.
Models incorporating coupled partial differential equations, describing
both morphological and chemical evolution, seem to offer the most
promise of explaining these observations. However, these models contain
many unknown parameters, which must satisfy specific conditions in
order to explain observed behavior. The lack of knowledge of these
parameters is therefore an important barrier to the comparison of
theory with experiment. Here, by adapting the recent theory of ``crater
functions'' to the case of binary materials, we develop a generic
framework in which many of the parameters of such models can be estimated
using the results of molecular dynamics simulations. 

As a demonstration, we apply our framework to the recent theory of
Bradley and Shipman, for the case of Ar-irradiated GaSb, in which
ordered patterns were first observed. In contrast to the requirements
therein that sputtered atoms form the dominant component of the collision
cascade, and that preferential redistribution play an important stabilizing
role, we find instead that the redistributed atoms dominate the collision
cascade, and that preferential redistribution appears negligible.
Hence, the actual estimated parameters for this system do not seem
to satisfy the requirements imposed by current theory, motivating
the consideration of other potential pattern-forming mechanisms.
\end{abstract}

\pacs{PACS numbers: 81.16.Rf, 79.20.Rf, 68.35.Ct, 02.70.Ns}

\maketitle

\section{Introduction}

The irradiation of flat surfaces by low-energy ions can lead to the
emergence of ordered arrays of nanoscale surface features \citep{facsko-etal-SCIENCE-1999},
and offers the potential to assist in the fabrication of nanostructured
devices with regular structure \citep{chan-chason-JAP-2007}. Initially
observed on a wide variety of target materials \citep{gago-etal-APL-2006,frost-etal-APA-2008,ziberi-etal-APL-2008},
careful experimental investigation has led to a growing consensus
that these structures only appear during the irradiation of binary
materials, or environments in which one or more materials is simultaneously
co-deposited during the irradiation process \citep{ozaydin-etal-APL-2005,ozaydin-etal-JVSTB-2008,ozaydin-ludwig-JPCM-2009,macko-etal-NanoTech-2010,zhang-etal-NJoP-2011,madi-aziz-ASS-2012}.
This consensus has motivated a number of theoretical treatments of
irradiated two-component materials, in which a coupled pair of partial
differential equations simultaneously track the evolution of both
the morphology and the concentration. Extending the earlier work of
Shenoy, Chan, and Chason \citep{shenoy-chan-chason-PRL-2007}, Bradley
and Shipman (BS) have recently introduced such a theory exhibiting
the first physically-grounded explanation for the strong ordering
of the observed nanostructures \citep{bradley-shipman-PRL-2010,shipman-bradley-PRB-2011}.
This result has sparked great interest, inspiring additional studies
on normal-incidence irradiation of two-component materials \citep{bradley-JAP-2012},
including a variant where the second material is not originally present,
but instead co-deposited simultaneously with the ion irradiation \citep{bradley-PRB-2011c},
and a similar framework for the study of ion-assisted deposition,
where both materials are deposited along with the ions, and the deposition
rate exceeds the sputtering rate \citep{abrasonis-morawetz-PRB-2012}.

Despite their successes, however, the models in this family face some
important open questions. In particular, they contain many competing
physical effects, represented by a large number of parameters whose
values are difficult to estimate experimentally, and are thus so far
undetermined. Although each theory seems reasonably able to explain
experimental observations for some region of parameter space, it is
unknown whether the various systems actually have parameters within
those regions. The lack of parameter estimates is therefore an important
barrier to the quantitative comparison of theory to experiment. In
this context, we will here develop a framework for the estimation
of several of these parameters by means of molecular dynamics simulations,
by extending the theory of ``crater functions'' \citep{norris-etal-2009-JPCM,norris-etal-NCOMM-2011}
to the case of binary materials.

We will begin in Section II with a necessary, brief review of the
key features of coupled-PDE theories, which have a common mathematical
structure and some generic stability properties. Then, in Section
III, we perform the first half of our task: we show how a generalization
of crater function theory, as in the case of pure materials, allows
the extraction of several terms of a continuum equation for the evolution
of the surface height, in terms of moments of the crater function
associated with simulations of single-ion impact. The second half
is to compare that equation with the corresponding equation in any
particular coupled-PDE model, by which one can in principle extract
several underlying parameters of the model, which in turn allow estimates
of the coefficients present in the PDEs.

Finally, in Section IV, we provide a lengthy demonstration of our
approach for the Bradley-Shipman theory itself \citep{bradley-shipman-PRL-2010,shipman-bradley-PRB-2011},
as applied to the long-studied $\mathrm{Ar}^{+}\to\mathrm{GaSb}$
system \citep{facsko-etal-SCIENCE-1999,le-roy-etal-PRB-2010-phase-separation,el-atwani-etal-JAP-2011}.
Estimating four important parameters for this theory using the the
framework developed in Section III, our principal findings are that
the destabilizing effect of sputtered atoms \citep{bradley-harper-JVST-1988}
continues to be overwhelmed by the stabilizing effect of redistributed
atoms \citep{carter-vishnyakov-PRB-1996,davidovitch-etal-PRB-2007},
and that the species-dependent redistribution seems to be negligible.
Both of these observations are contrary to the requirements for ordered
pattern formation imposed by the Bradley-Shipman model, and suggest
that additional or alternate physical mechanisms may be necessary
to explain this phenomena.

\section{Review of Existing Theory: Models and Stability Analysis}

We begin by briefly summarizing a recent class of models including
contributions from Shenoy, Chan, and Chason \citep{shenoy-chan-chason-PRL-2007},
Bradley and Shipman \citep{bradley-shipman-PRL-2010,shipman-bradley-PRB-2011},
Bradley \citep{bradley-PRB-2011c,bradley-JAP-2012}, and Abrasonis
and Morowetz \citep{abrasonis-morawetz-PRB-2012} for various irradiation
regimes involving two target species. Because extensive analysis of
this class of models exists elsewhere, both in the just-cited works,
and in general \citep{cross-greenside-2009-book}, we will be as brief
as possible. However, because our aim is to test specific aspects
of these theories, we must include those results that are directly
relevant to the molecular dynamics simulations we perform later.

\subsection{Models}

The models cited above all consider the normal-incidence irradiation
of an initially flat target which contains two components in one of
three ways: (a) the target itself is a binary material \citep{bradley-shipman-PRL-2010,shipman-bradley-PRB-2011,bradley-JAP-2012},
(b) the target is pure, but a second material is co-deposited during
sputtering \citep{bradley-PRB-2011c}, or (c) two materials are co-deposited
at a rate that exceeds the net sputter rate \citep{abrasonis-morawetz-PRB-2012}
(termed Ion Beam Assisted Deposition or IBAD). In any of these cases,
one tracks the evolution of a height field $z=h\left(x,y,t\right)$
describing the irradiated surface, and concentration fields $c_{A}\left(x,y,t\right)$
and $c_{B}\left(x,y,t\right)$ of two components $A$ and $B$. Under
the effects of preferential sputtering, a steady state is reached
in which the material is receding (or advancing, for IBAD) with with
constant velocity $v_{0}$ and constant surface concentrations $c_{A,0}$
and $c_{B,0}$ of $A$ and $B$ atoms, respectively. The stability
of this steady state is then investigated by introducing perturbations
of the form 
\begin{equation}
\begin{aligned}h & =-v_{0}t+u\left(x,y,t\right)\\
c_{A} & =c_{A,0}+\phi\left(x,y,t\right)\\
c_{B} & =c_{B,0}+\left(1-\phi\left(x,y,t\right)\right)
\end{aligned}
\label{eq: perturbations}
\end{equation}
where $u\left(x,y,t\right)$ describes a small perturbation to the
height field, and $\phi\left(x,y,t\right)$ describes a small perturbation
to the concentration field of species $A$. After significant modeling
of various physical effects during ion irradiation, each of Refs.\citep{bradley-shipman-PRL-2010,shipman-bradley-PRB-2011,bradley-JAP-2012,bradley-PRB-2011c,abrasonis-morawetz-PRB-2012}
obtains linearized equations for the evolution of $u$ and $\phi$
of the form

\begin{eqnarray}
\frac{\partial u}{\partial t} & = & -A\phi+B\nabla^{2}\phi+C\nabla^{2}u-D\nabla^{4}u\label{eqn: linear-height-evolution-BH}\\
\frac{\partial\phi}{\partial t} & = & -A^{\prime}\phi+B^{\prime}\nabla^{2}\phi+C^{\prime}\nabla^{2}u-D^{\prime}\nabla^{4}u.\label{eqn: linear-concentration-evolution-BH}
\end{eqnarray}
The exact meaning of the coefficients varies among the cited models,
but in general the following interpretations may be provided. In Eq.~(\ref{eqn: linear-height-evolution-BH})
describing the height field: the term $-A\phi$ describes the concentration
dependence of the net erosion rate, so that if $A>0$ then increasing
the concentration of species $A$ increases the sputter yield \citep{shenoy-chan-chason-PRL-2007};
the term $B\nabla^{2}\phi$ describes a \emph{net} mass flux along
concentration gradients caused by unequal diffusivities between the
components; the term $C\nabla^{2}u$ captures the net effects of both
curvature dependent sputtering \citep{bradley-harper-JVST-1988} and
mass redistribution \citep{carter-vishnyakov-PRB-1996} (and also
stress \citep{castro-cuerno-ASS-2012,norris-PRB-2012-linear-viscous,castro-etal-PRB-2012},
which will however be neglected here); the term $-D\nabla^{4}u$ (with
$D>0$ by definition) describes many kinds of surface relaxation including
surface diffusion \citep{mullins-JAP-1959,bradley-harper-JVST-1988}
and surface-confined viscous flow \citep{orchard-ASR-1962,umbach-etal-PRL-2001},
and regularizes the height dependence in the case that $C$ is negative.
In Eq.(\ref{eqn: linear-concentration-evolution-BH}) describing the
concentration field: the term $-A^{\prime}\phi$ (with $A^{\prime}>0$
by definition) describes the continual resupply of material into the
irradiated film at some reference concentration of either the bulk
(in the regime of erosion) or the vapor (in the regime of growth),
which serves to damp perturbations away from this concentration; the
term $B^{\prime}\nabla^{2}\phi$ describes simple Fickian diffusion;
the term $C^{\prime}\nabla^{2}u$ describes species-dependent redistribution
\citep{bradley-shipman-PRL-2010}, and the term $-D^{\prime}\nabla^{4}u$
would describe preferential migration away from regions of high curvature,
due to differing atomic mobilities.

\subsection{Generic Stability Analysis\label{sub: Stability-Analysis}}

To determine the presence or absence of an instability, is is sufficient
to consider a one-dimensional perturbation, because of the isotropic
nature of normal-incidence irradiation. Without loss of generality,
we may orient this instability in the $x$-coordinate direction, giving

\begin{equation}
\left[\begin{array}{c}
u_{\phantom{1}}\left(x,t\right)\\
\phi_{\phantom{1}}\left(x,t\right)
\end{array}\right]=\left[\begin{array}{c}
u_{1}\\
\phi_{1}
\end{array}\right]e^{ikx+\sigma t},\label{eqn: infinitesimal-perturbation}
\end{equation}
 where $k$ is the wavenumber of the perturbation, $\sigma\left(k\right)$
the (wavenumber-dependent) growth rate of that perturbation, and $u_{1}$
and $\phi_{1}$ are undetermined constants describing the relative
phases and magnitudes of the height and concentration modulations.
Inserting the ansatz (\ref{eqn: infinitesimal-perturbation}) into
the linearized equations (\ref{eqn: linear-height-evolution-BH})-(\ref{eqn: linear-concentration-evolution-BH}),
we obtain, in matrix form,
\begin{equation}
\left[\begin{array}{cc}
\sigma+Ck^{2}+Dk^{4} & A+Bk^{2}\\
C^{\prime}k^{2}+D^{\prime}k^{4} & \sigma+A^{\prime}+B^{\prime}k^{2}
\end{array}\right]\left[\begin{array}{c}
u_{1}\\
\phi_{1}
\end{array}\right]=\mathbf{0}.\label{eqn: stability-matrix}
\end{equation}
For a solution $\left[u_{1},\,\phi_{1}\right]^{T}$ of this equation,
the determinant of the matrix must be zero, giving a the dispersion
relation $\sigma\left(k\right)$ in the quadratic form 
\begin{equation}
\sigma^{2}+\tau\sigma+\Delta=0,\label{eqn: quadratic-equation}
\end{equation}
where $\tau\left(k\right)$ and $\Delta\left(k\right)$ are respectively
the trace and determinant of (\ref{eqn: stability-matrix}) when $\sigma=0$:
\begin{eqnarray}
\tau\left(k\right) & = & A^{\prime}+\left(C+B^{\prime}\right)k^{2}+Dk^{4}\label{eqn: tau}\\
\Delta\left(k\right) & = & \left(CA^{\prime}-AC^{\prime}\right)k^{2}+\left(CB^{\prime}-C^{\prime}B+DA^{\prime}-D^{\prime}A\right)k^{4}+\left(DB^{\prime}-D^{\prime}B\right)k^{6}.\label{eqn: Delta}
\end{eqnarray}
For the purpose of identifying stability properties, it is sufficient
to consider only the root associated with the '+' sign in the quadratic
formula which, following common practice, is denoted $\sigma_{+}\left(k\right)$:
\begin{equation}
\sigma_{+}\left(k\right)=\frac{1}{2}\left(-\tau+\sqrt{\tau^{2}-4\Delta}\right).\label{eqn: quadratic-solution-of-sigma}
\end{equation}
This gives the growth rate of the faster-growing solution when the
discriminant is positive.

\subsection{Discussion: Pattern Formation and Coefficient Values\label{sub: finite-wavelength-requirements}}

\begin{figure}
\centering{}\includegraphics[width=2.5in]{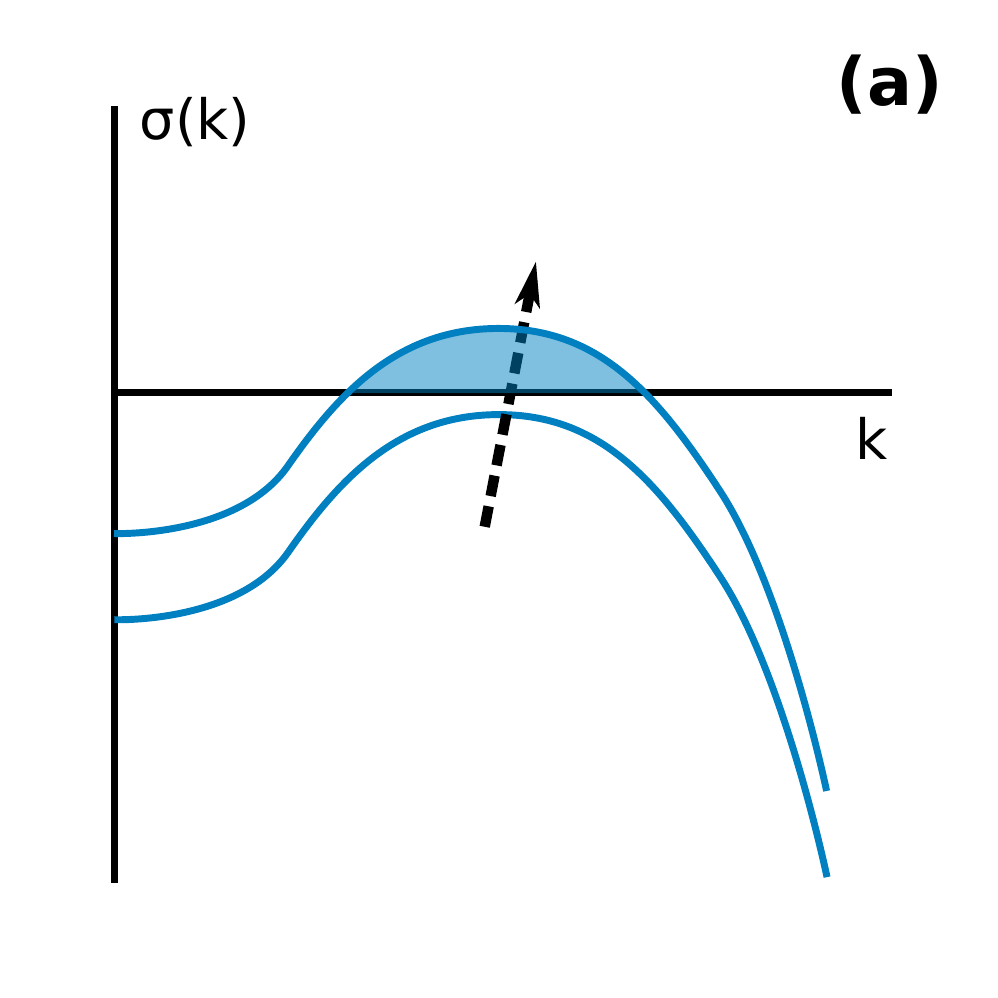}\includegraphics[width=2.5in]{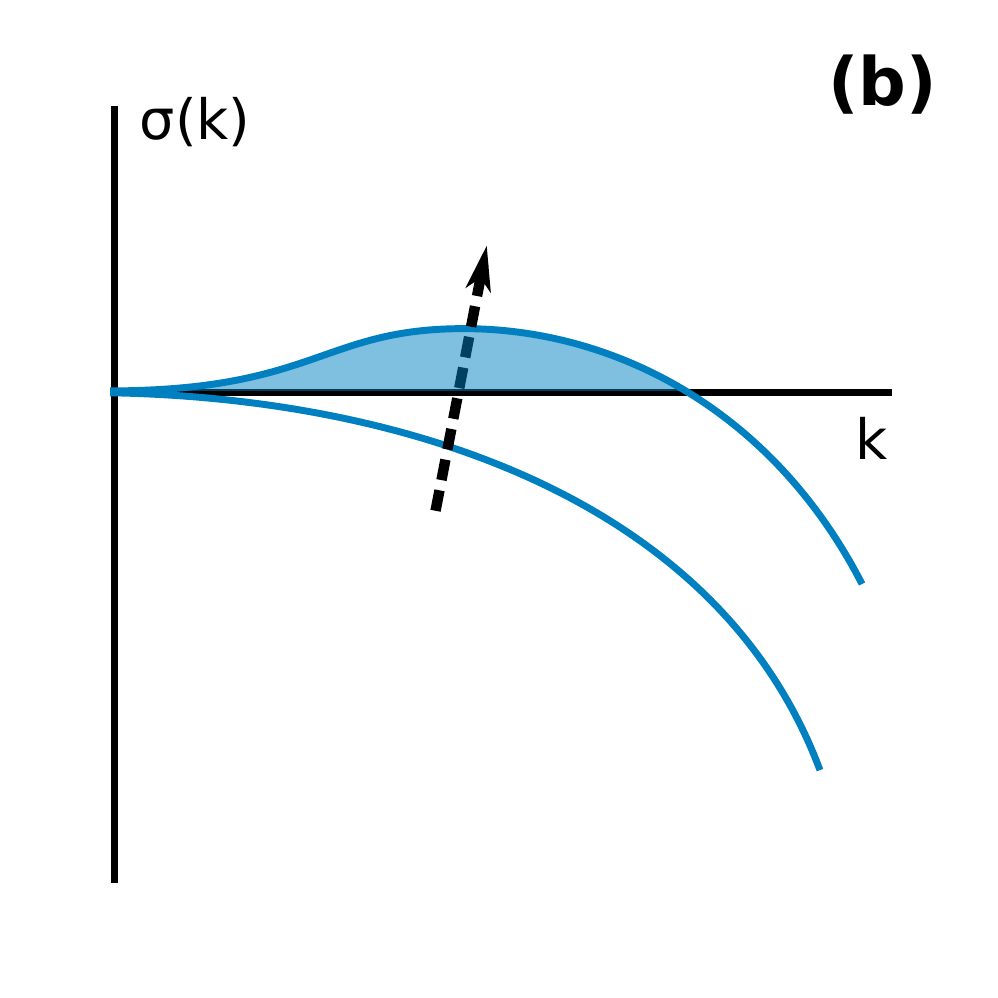}\caption{(a) A finite-wavelength instability, in which a narrow band of unstable
wavenumbers emerges, bounded above and below by stable wavenumbers.
Systems with linear instabilities of this kind generically exhibit
some kind of ordered pattern. (b) A long-wavelength instability, in
which all wavenumbers below a critical value are unstable. Systems
with linear instabilities of this kind generically exhibit roughening
and coarsening.\label{fig: instability-types}}
\end{figure}
The expected behavior of an initially-flat surface is determined by
the dispersion relation $\sigma_{+}\left(k\right)$. Wavenumbers $k$
for which $\sigma_{+}\left(k\right)<0$ are stable and decay over
time, while those $k$ for which $\sigma_{+}\left(k\right)>0$ are
unstable and grow over time.  Provided at least some wavenumbers
are unstable, then some kind of structure emerges on the irradiated
surface. However, the nature and evolution of this structure depends
strongly on the nature of $\sigma_{+}\left(k\right)$, as extensively
discussed in Refs.~\citep{cross-greenside-2009-book}. In particular,
if the unstable wavenumbers exist only in a narrow band, surrounded
on both sides by stable wavenumbers, then the limited range of unstable
wavelengths drives the system into a relatively well-ordered configuration
that preserves is characteristic size over time (Figure~\ref{fig: instability-types}a).
On the other hand, if all wavenumbers below a critical value are unstable,
then the system tends to roughen and coarsen over time, as the slower-growing
but energetically-favorable longer wavelengths take over from the
initial instability (Figure~\ref{fig: instability-types}b).

Based on these considerations, it is clear that a key determinant
of the observed pattern is the stability of the longwaves: if a system
is observed to exhibit well-ordered patterns in the linear regime,
it must be true that the longwaves are linearly stable. From Eqs.~(\ref{eqn: tau})-(\ref{eqn: quadratic-solution-of-sigma})
in the limit as $k\to0$, we can easily find that for small values
of $k$
\begin{equation}
\lim_{k\to0}\sigma_{+}\left(k\right)\approx-\frac{1}{A^{\prime}}\left(A^{\prime}C-C^{\prime}A\right)k^{2}+\mathcal{O}\left(k^{4}\right).\label{eqn: small-k-limit}
\end{equation}
Hence, any system with well-ordered patterns must satisfy the requirement

\begin{equation}
A^{\prime}C-C^{\prime}A>0;\label{eqn: finite-wavelength-requirement}
\end{equation}
this provides a tangible and concrete test for the comparison of theory
with experiment. We see that this important predictor of qualitative
system behavior depends on no less than four of the unknown theoretical
parameters, which highlights the importance of finding ways to produce
estimates of these coefficients. More intriguingly, we see from the
definitions above that each of these parameters is associated in some
way with the effect of ion impacts, which suggests the possibility
of estimating these parameters via molecular dynamics.

\section{Coefficient Estimation via Molecular Dynamics: Crater Functions for
Compound Materials\label{sec: Coefficient-Estimation}}

We now develop a framework in which the results of MD simulation may
be connected to the coefficients of Eqs.(\ref{eqn: linear-height-evolution-BH})-(\ref{eqn: linear-concentration-evolution-BH}).
To this end, we will generalize the ``crater function'' theory for
surface morphology evolution \citep{norris-etal-2009-JPCM,norris-etal-NCOMM-2011}
- which extracts certain PDE terms directly from molecular dynamics
simulations - to the case of a multi-component material. Because the
generalization is straightforward, we will rely heavily on the referenced
works, and present only the differences in the present approach.

We begin by taking the ``crater function'' $\Delta h$$\left(x,y;\theta\right)$,
describing the average change in surface morphology to a flat surface
due to the impact of a single ion at the origin, and splitting it
into four parts:
\begin{equation}
\Delta h\left(x,y;\theta,c_{A},c_{B}\right)=\Delta h_{A}^{\mathrm{eros.}}+\Delta h_{B}^{\mathrm{eros.}}+\Delta h_{A}^{\mathrm{redist.}}+\Delta h_{B}^{\mathrm{redist.}}.\label{eqn: compound-crater-functions}
\end{equation}
This merely states that the change in surface topography due to a
single ion impact is the superposition of effects due to both eroded
and redistributed atoms, of both species A and species B. A central
benefit of this approach is that any subsequent analytical treatment
of $\Delta h$ treats all these effects simultaneously. Therefore,
because the multi-scale analysis of Ref.\citep{norris-etal-2009-JPCM}
(and a simplified approximation for flat targets in Ref.\citep{norris-etal-NCOMM-2011})
has already been performed for a general $\Delta h$, all the resulting
PDE terms therein can be similarly decompose\ref{eqn: compound-crater-functions}d
into contributions of the same four types.

To estimate the components in Eq.(\ref{eqn: compound-crater-functions})
from a molecular dynamics simulation, we follow Ref.\citep{norris-etal-2009-JPCM}
by assuming that all displacements within the bulk are immediately
projected to the surface to cause a change in the surface profile,
which can then be approximated by placing delta functions at the initial
and final position of each atom. Using Eqs.(46) and (48) of Ref.\citep{norris-etal-2009-JPCM}
as a template, we define 
\begin{equation}
\begin{aligned}\Delta h_{A}^{\mathrm{eros.}} & =-\Omega_{A}\sum_{i}\delta\left(\mathbf{x}_{i}^{I}\right)\\
\Delta h_{B}^{\mathrm{eros.}} & =-\Omega_{B}\sum_{j}\delta\left(\mathbf{x}_{j}^{I}\right)\\
\Delta h_{A}^{\mathrm{redist.}} & =\phantom{-}\Omega_{A}\sum_{k}\delta\left(\mathbf{x}_{k}^{F}\right)-\delta\left(\mathbf{x}_{k}^{I}\right)\\
\Delta h_{B}^{\mathrm{redist.}} & =\phantom{-}\Omega_{B}\sum_{l}\delta\left(\mathbf{x}_{l}^{F}\right)-\delta\left(\mathbf{x}_{l}^{I}\right)
\end{aligned}
\label{eq: crater-function-contributions}
\end{equation}
where the $\Omega$ denote atomic volumes, $i,j$ denote sputtered
$A$,$B$ atoms, $k,l$ denote displaced $A$,$B$ atoms, and $\mathbf{x}^{I}$
and $\mathbf{x}^{F}$ denote initial and final target atom positions
associated with an ion impact. 

A central result of Refs.\citep{norris-etal-2009-JPCM,norris-etal-NCOMM-2011}
is that morphology evolution depends only on the moments of the crater
function (\ref{eqn: compound-crater-functions}). When the crater
function is integrated to obtain these moments, each moment automatically
contains associated contributions of each type. Following Eqs.(47)
and (49) of Ref.\citep{norris-etal-2009-JPCM} by directly integrating
the components in Eq.\ref{eq: crater-function-contributions}, we
obtain the moments via
\begin{equation}
\begin{aligned}M_{A}^{\left(0\right)} & =\int\Delta h_{A}^{\mathrm{eros.}}dA & =-\Omega_{A}\sum_{i}1\\
M_{B}^{\left(0\right)} & =\int\Delta h_{B}^{\mathrm{eros.}}dA & =-\Omega_{B}\sum_{j}1\\
M_{A,\mathrm{eros.}}^{\left(1\right)} & =\int\mathbf{x}\Delta h_{A}^{\mathrm{eros.}}dA & =-\Omega_{A}\sum_{i}\mathbf{x}_{i}^{\mathrm{I}}\\
M_{B,\mathrm{eros.}}^{\left(1\right)} & =\int\mathbf{x}\Delta h_{B}^{\mathrm{eros.}}dA & =-\Omega_{B}\sum_{j}\mathbf{x}_{j}^{\mathrm{I}}\\
M_{A,\mathrm{redist.}}^{\left(1\right)} & =\int\mathbf{x}\Delta h_{A}^{\mathrm{redist.}}dA & =\phantom{-}\Omega_{A}\sum_{k}\left(\mathbf{x}_{k}^{F}-\mathbf{x}_{k}^{\mathrm{I}}\right)\\
M_{B,\mathrm{redist.}}^{\left(1\right)} & =\int\mathbf{x}\Delta h_{B}^{\mathrm{redist.}}dA & =\phantom{-}\Omega_{B}\sum_{l}\left(\mathbf{x}_{l}^{F}-\mathbf{x}_{l}^{\mathrm{I}}\right)
\end{aligned}
.\label{eqn: moments-from-positions}
\end{equation}
(note that the zeroth moment has no redistributive part). Although
additional moments exist and contribute, in principle, to the dynamics
of the surface, no theory to date incorporates these effects; as our
present goal is to estimate the coefficients of existing theories,
we will not consider higher moments here.

Finally, following a multi-scale analysis, these contributions to
the moments produce associated terms in the governing equation for
the height field
\begin{equation}
\begin{aligned}h_{t} & \approx\left(Y^{A}+Y^{B}\right)\\
 & +\left(S_{X}^{A,\mathrm{eros.}}+S_{X}^{B,\mathrm{eros.}}+S_{X}^{A,\mathrm{redist.}}+S_{X}^{B,\mathrm{redist.}}\right)h_{xx}\\
 & +\left(S_{Y}^{A,\mathrm{eros.}}+S_{Y}^{B,\mathrm{eros.}}+S_{Y}^{A,\mathrm{redist.}}+S_{Y}^{B,\mathrm{redist.}}\right)h_{yy}
\end{aligned}
\label{eqn: pure-h-equation}
\end{equation}
with definitions given by Eq.(4) of Ref.\citep{norris-etal-NCOMM-2011}
\begin{equation}
\begin{aligned}Y^{Z}\left(\theta,c_{i}\right) & =\phantom{\frac{\partial}{\partial\theta}}\left[I_{0}\cos\left(\theta\right)M_{Z}^{\left(0\right)}\left(\theta,c_{i}\right)\right]\\
S_{X}^{Z\mathrm{,type}}\left(\theta,c_{i}\right) & =\frac{\partial}{\partial\theta}\left[I_{0}\cos\left(\theta\right)M_{Z,\mathrm{type}}^{\left(1\right)}\left(\theta,c_{i}\right)\right]\\
S_{Y}^{Z,\mathrm{type}}\left(\theta,c_{i}\right) & =\phantom{\frac{\partial}{\partial\theta}}\left[I_{0}\cos\left(\theta\right)\cot\left(\theta\right)M_{Z,\mathrm{type}}^{\left(1\right)}\left(\theta,c_{i}\right)\right]
\end{aligned}
,\label{eqn: moments-to-coeffs}
\end{equation}
where $Z$ is either $A$ or $B$, 'type' is either 'eros.' or 'redist.',
and $I_{0}$ is the flux through a plane perpendicular to the beam
(see Section~\ref{sub:Limitations} for further comments on these
formulas). 

Equation~(\ref{eqn: pure-h-equation}) completely describes the response
of the surface to the effects of sputtered atoms (via the base yield
terms $Y^{Z}$ and curvature-dependent yield terms $S^{Z,\mathrm{eros.}}$),
and also to the effects of redistributed atoms (via the terms $S^{Z,\mathrm{redist.}}$).
Although it applies only to the height field, and our aim here is
to estimate the behavior of multi-component theories, we now have
sufficient information to estimate several parameters within these
theories. By carefully examining the equation for the hight field
in any given two-component theory, we can identify the associated
terms, and equate the coefficients in Eq.(\ref{eqn: pure-h-equation})
with \emph{underlying parameter} \emph{groups} in the theory. These
underlying parameter groups, in turn, will allow us to estimate several
of the coefficients in equations (\ref{eqn: linear-height-evolution-BH})-(\ref{eqn: linear-concentration-evolution-BH}).
Although this approach defies a single mathematical notation due to
differences between theories, it is straightforward, and we shall
now offer a lengthy demonstration for an important example problem.

\section{An Example Application: Irradiated Binary Alloys\label{sec: application-to-GaSb}}

We now present an application of the general method described above
to the regime of irradiated binary alloys. The most detailed coupled-PDE
theory for this system is that of Bradley and Shipman \citep{bradley-shipman-PRL-2010,shipman-bradley-PRB-2011},
which generalizes earlier work of Shenoy, Chan, and Chason \citep{shenoy-chan-chason-PRL-2007}
to include the effect of both net and preferential redistribution
of atoms that are not sputtered away from the surface -- i.e., the
so-called Carter-Vishnyakov effect \citep{carter-vishnyakov-PRB-1996}.
Although it is a general framework applicable to the irradiation of
any binary material, the work of Bradley and Shipman had primarily
in mind the irradiation of GaSb, where ordered hexagonal dot structures
were famously first observed \citep{facsko-etal-SCIENCE-1999}. We
therefore illustrate how one may use molecular dynamics simulations
to estimate the value of parameters in the theory for GaSb. (Note:
when referring to specific equations within the BS theory, we refer
to the longer and more detailed Ref.~\citep{shipman-bradley-PRB-2011},
despite the earlier publication of Ref.~\citep{bradley-shipman-PRL-2010}).

\subsection{The Bradley-Shipman Theory and Connection with Crater Functions}

The Bradley-Shipman theory starts from Eqs.(\ref{eqn: linear-height-evolution-BH})-(\ref{eqn: linear-concentration-evolution-BH}),
and applies the simplifying assumption that one can neglect the terms
$B\nabla^{2}c$ in Eq.(\ref{eqn: linear-height-evolution-BH}) and
$D^{\prime}\nabla^{4}h$ in Eq.(\ref{eqn: linear-concentration-evolution-BH})
due to equal mobilities of target atoms. We deviate slightly from
Ref.\citep{shipman-bradley-PRB-2011} by assuming that species $A$
is preferentially sputtered (rather than species $B$), and define
coefficients in such a way that zeroth- and fourth-order terms in
Eqs.(\ref{eqn: linear-height-evolution-BH})-(\ref{eqn: linear-concentration-evolution-BH})
have negative sign, whereas second-order terms have positive sign.
The coefficients in Eq.(\ref{eqn: linear-height-evolution-BH}) for
the height evolution are then defined via

\begin{equation}
\begin{aligned}A & =P_{0}\Omega\left[\Lambda_{A}^{\prime}\left(c_{A,0}\right)-\Lambda_{B}^{\prime}\left(c_{B,0}\right)\right]\\
C & =\Omega\left[\left(\mu_{A}\left(c_{A,0}\right)+\mu_{B}\left(c_{B,0}\right)\right)-\alpha\left(\Lambda_{A}\left(c_{A,0}\right)+\Lambda_{B}\left(c_{B,0}\right)\right)\right]\\
D & =\left[c_{A,0}D_{A}+c_{B,0}D_{B}\right]\frac{n_{s}\Omega^{2}\gamma_{s}}{k_{B}T}>0
\end{aligned}
\label{eq: ACD}
\end{equation}
and in Eq.(\ref{eqn: linear-concentration-evolution-BH}) for the
concentration evolution, via

\begin{equation}
\begin{aligned}A^{\prime} & =\frac{P_{0}\Omega}{\Delta}\left[c_{B,b}\Lambda_{A}^{\prime}\left(c_{A,0}\right)+c_{A,b}\Lambda_{B}^{\prime}\left(c_{B,0}\right)\right]>0\\
B^{\prime} & =\frac{n_{s}\Omega}{\Delta}\left[c_{B,b}D_{A}+c_{A,b}D_{B}\right]>0\\
C^{\prime} & =\frac{\Omega}{\Delta}\left[c_{B,b}\mu_{A}\left(c_{A,0}\right)-c_{A,b}\mu_{B}\left(c_{B,0}\right)\right]
\end{aligned}
.\label{eq: alpha-beta-gamma}
\end{equation}
Here $P_{0}$ is the power deposited by the ions per unit surface
area on a flat surface, $\Omega$ is the atomic volume (taken to be
the same for both species), $n_{s}$ is the total number of mobile
surface atoms per unit area on the surface, $\gamma_{s}$ is the surface
energy, $k_{B}$ is Boltzmann's constant, $T$ is the temperature,
$\Delta$ is the amorphous film thickness, $c_{A,0}$ and $c_{B,0}$
are the concentration of $A$ and $B$ atoms in the film at steady
state, $c_{A,b}$ and $c_{B,b}$ are the concentration of $A$ and
$B$ atoms in the bulk, $D_{A}$ and $D_{B}$ are the diffusivities
of $A$ and $B$ atoms, $\Lambda_{A}\left(c_{A}\right)$ and $\Lambda_{B}\left(c_{B}\right)$
are proportionality constants linking the deposited power to the sputtering
rate, and $\mu_{A}\left(c_{A}\right)$ and $\mu_{B}\left(c_{B}\right)$
are proportionality constants describing preferential redistribution
of material, an effect first proposed in Refs.\citep{bradley-shipman-PRL-2010,shipman-bradley-PRB-2011}
and playing a critical role therein.

To connect with the results of Section~\ref{sec: Coefficient-Estimation},
the key equation in the Bradley-Shipman theory is Eq.(7) of Ref.\citep{shipman-bradley-PRB-2011},
with additional definitions provided by Eqs.(3)-(6) therein. Combining
those four equations, one obtains an evolution equation for the height
field of 

\begin{equation}
\frac{\partial h}{\partial t}=-\Omega P_{0}\left(\Lambda_{A,0}+\Lambda_{B,0}\right)+\Omega\left(-\alpha\Lambda_{A,0}-\alpha\Lambda_{B,0}+\mu_{A}+\mu_{B}\right)\nabla^{2}h+\dots,\label{eq: BH-height-only}
\end{equation}
where all terms directly due to ion-bombardment are present, and terms
related to diffusive processes are omitted. Upon comparison of Eq.(\ref{eq: BH-height-only})
with Eq.(\ref{eqn: pure-h-equation}), it can easily be seen that
there is a one-to-one correspondence between the coefficients of the
two equations, and the physical effect underlying each. Although these
equations were obtained by different means, they describe precisely
the same physical mechanisms. We therefore equate them in a term-by-term
manner for the case of normal incidence ($\theta=0$), obtaining

\begin{equation}
\begin{aligned}\Omega_{A}P_{0}\Lambda_{A,0} & =-Y^{A}\\
\Omega_{B}P_{0}\Lambda_{B,0} & =-Y^{B}\\
\Omega_{A}\alpha\Lambda_{A,0} & =-S_{X,Y}^{A,\mathrm{eros.}}\\
\Omega_{B}\alpha\Lambda_{B,0} & =-S_{X,Y}^{B,\mathrm{eros.}}\\
\Omega_{A}\mu_{A,0} & =\phantom{-}S_{X,Y}^{A,\mathrm{redist.}}\\
\Omega_{B}\mu_{B,0} & =\phantom{-}S_{X,Y}^{B,\mathrm{redist.}}
\end{aligned}
\label{eqn: craters-to-BS-theory}
\end{equation}
Finally, combining the relationships (\ref{eqn: craters-to-BS-theory})
with the definitions for $\left\{ A,C,A^{\prime},C^{\prime}\right\} $
in Eqs.(\ref{eq: ACD})-(\ref{eq: alpha-beta-gamma}), we can then
estimate these parameters as follows 
\begin{equation}
\begin{aligned}A & =-\left(\left[Y^{A}\right]^{\prime}-\left[Y^{B}\right]^{\prime}\right)\\
C & =\left(c_{A,0}S_{X,Y}^{A,\mathrm{eros.}}+c_{B,0}S_{X,Y}^{B,\mathrm{eros.}}+c_{A,0}S_{X,Y}^{A,\mathrm{redist.}}+c_{B,0}S_{X,Y}^{B,\mathrm{redist.}}\right)\\
A^{\prime} & =-\frac{1}{\Delta}\left(c_{A,b}\left[Y^{B}\right]^{\prime}+c_{B,b}\left[Y^{A}\right]^{\prime}\right)>0\\
C^{\prime} & =\frac{1}{\Delta}\left(c_{B,b}c_{A,0}S_{X,Y}^{A,\mathrm{redist.}}-c_{A,b}c_{B,0}S_{X,Y}^{B,\mathrm{redist.}}\right)
\end{aligned}
.\label{eq: craters-to-BS-conversion}
\end{equation}
We see that fully four out of six of the BS parameters are accessible
to MD studies. As we anticipated in Section \ref{sub: finite-wavelength-requirements},
these happen to be precisely the parameters that govern the behavior
of the longest wave perturbations to the surface height, via the finite-wavelength
requirement in Eq.(\ref{eqn: finite-wavelength-requirement}). Thus,
the results of MD simulation may be used to predict whether an irradiated
binary system should have long-wave or finite-wavelength instabilities,
a surprising result that allows the prediction of a key distinguishing
characteristic of pattern-forming systems.

\subsection{Simulation Environment and Initial Results\label{sub: environment-and-results}}

Using molecular dynamics simulations, we have performed simulations
of GaSb irradiated by Ar ions at 250 eV. The simulations were carried
out with the PARCAS code \citep{nordlund-parcas}, following similar
general principles for surface irradiation simulations as presented
earlier in Refs. \citep{tarus-etal-PRB-1998,ghaly-nordlund-averback-1999-PMA}.
The same simulation approach has earlier been shown to give reasonable
agreement with experimental sputtering yields in Si \citep{samela-etal-2007-NIMB}.
Amorphous GaSb was created by annealing a cubic piece of initially
crystalline GaSb, consisting of 1.1 x 10\textsuperscript{4} atoms.
In these simulations periodic boundary conditions were used. Copies
of the obtained amorphous GaSb were joined together (3 in the $x$-
and $y$-directions, and 2 in the $z$-direction) to form a target
consisting of nearly 2.0 x 10\textsuperscript{5} atoms. Finally, the
target was cleaved in the $z$-direction, and the resulting free surface
was relaxed at 0 K, with periodicity only in the $x$- and $y$-directions. 

The inter-atomic potentials used for Ga-Sb interactions was that of
Powell et al \citep{powell-etal-PRB-2007}. The (rarely occurring)
pure Ga interactions were the same as those in the GaAs potential
developed by us \citep{albe-etal-PRB-2002}. To obtain the Sb potential,
we note that As and Sb have the same crystal structure. Hence, the
As potential parameter $D_{e}$ \citep{albe-etal-PRB-2002} was rescaled
with the difference in cohesive energies of the two elements, and
the parameter $r_{0}$ with the difference in lattice constants. This
gave a cohesive energy of 2.72 eV/atom and a density of $7.28\,\mathrm{g}/\mathrm{cm}^{3}$
for pure Sb, in reasonable agreement with the experimental values
of 2.72 eV/atom and $6.74\,\mathrm{g}/\mathrm{cm}^{3}$. We emphasize
that since regions of pure Ga or Sb should not form in the simulated
system, the results are not expected to be sensitive to small inaccuracies
in the description of the pure elements.

For each different irradiation angle, we simulated 300 independent
Ar ion impacts onto the target just constructed. Each ion hit the
target at a random location, and for off normal incidence, arrived
from a random azimuthal angle. Periodic boundary conditions were used
in the x and y directions, with a thermostat dynamically applied to
vertical and horizontal strips of material halfway across the cell
from the impact position, so as to form cooled walls within which
the impact was centered. Together with a thermostat layer at the bottom
of the cell, these served to remove energy from the system over time.
The simulation time of each impact was 250 ps, which is rather long,
but turned out to be necessary for the post-impact atomic movements
to seize. 

As discussed in Ref.\citep{norris-etal-NCOMM-2011}, two sources of
noise were then filtered from the data before the moments were measured.
First, despite the careful preparation of the target, some residual
stresses remain, leading to displacements during the simulation that
are due to the pre-existing state of the target, rather than to the
impact. Second, the periodic nature of the cell allows small coherent
shears to occur, on the scale of the entire cell. These are both filtered
by constructing an annulus around the impact cite, and calculating
both the average lab-frame displacement of each atom, as well as the
average shear in a co-ordinate system aligned with the incoming ion
direction. Both averages were then subtracted from the displacement
field, after which moments were calculated according to Eq.(\ref{eqn: moments-from-positions}).

\begin{center}
\begin{figure}
\begin{centering}
\includegraphics[width=2.15in]{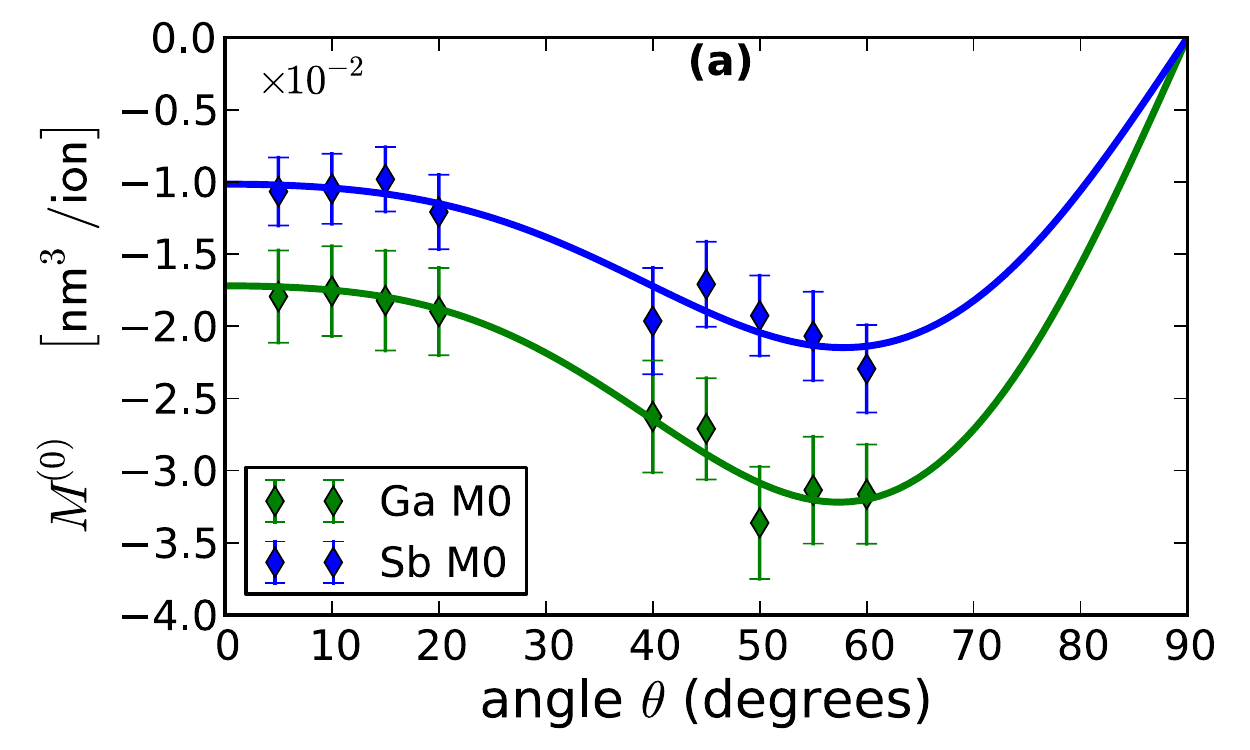}\includegraphics[width=2.15in]{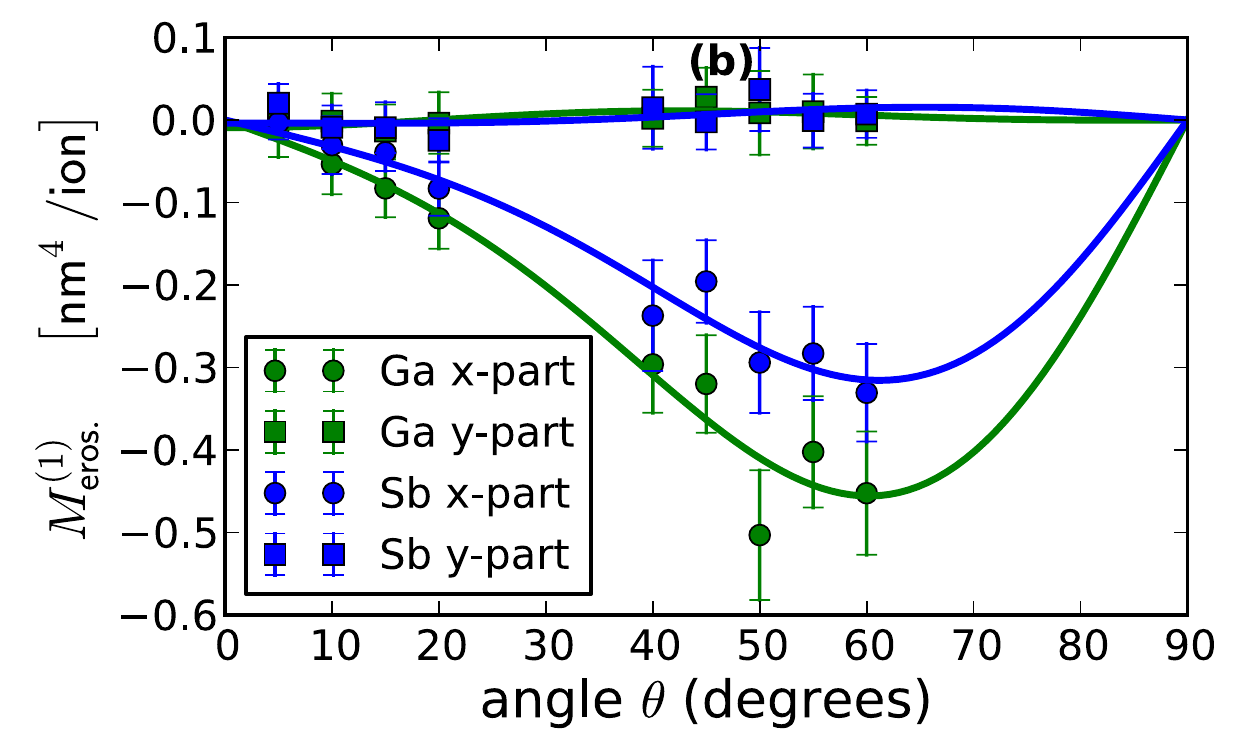}\includegraphics[width=2.15in]{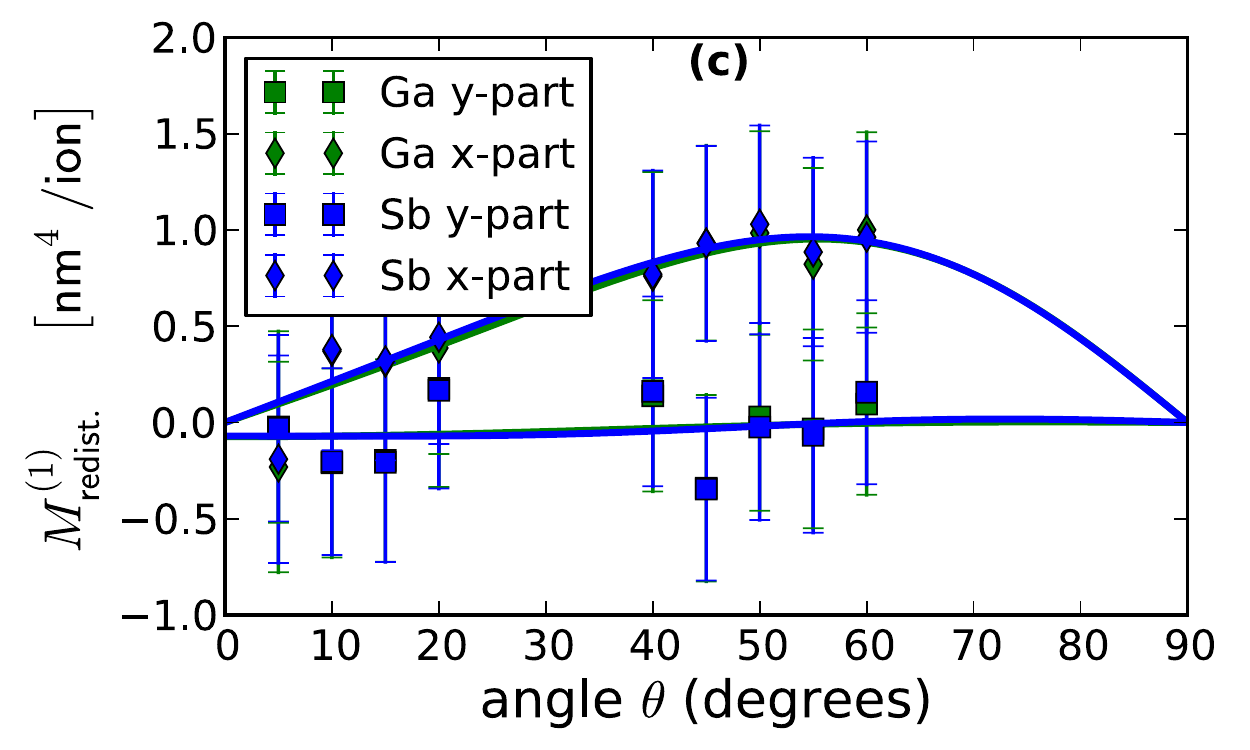}
\par\end{centering}

\begin{centering}
\includegraphics[width=2.15in]{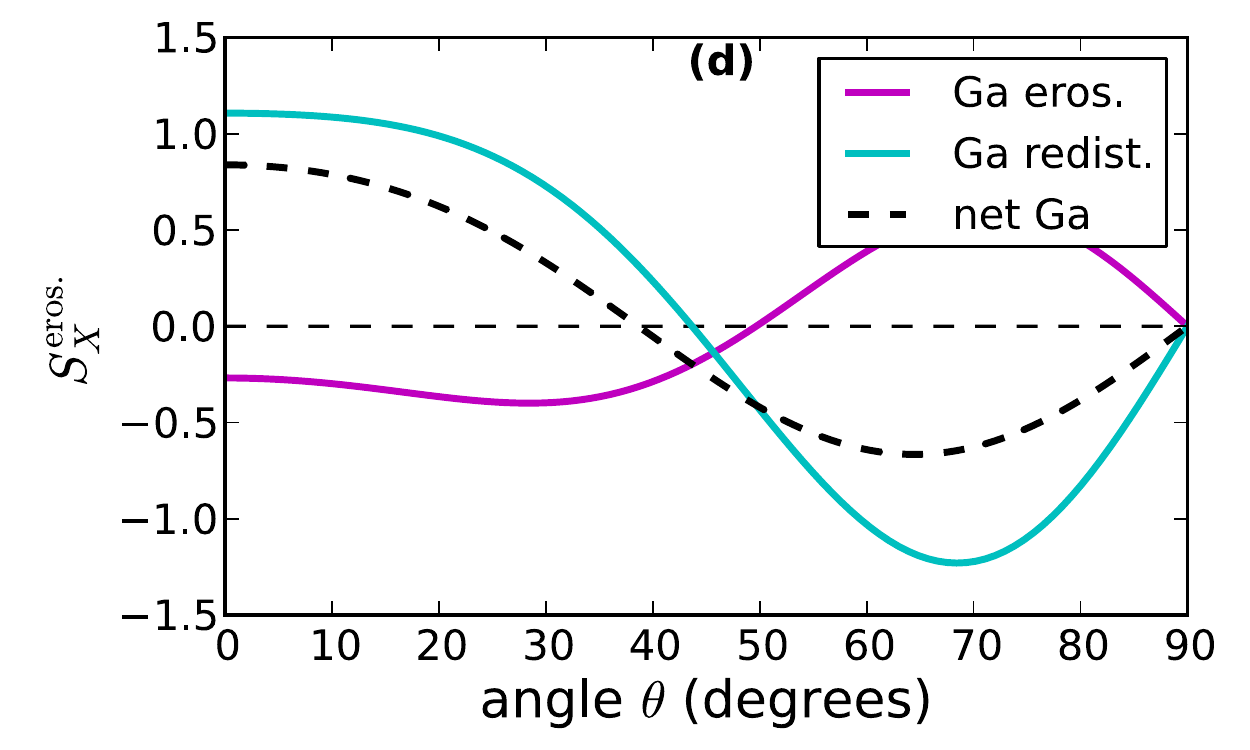}\includegraphics[width=2.15in]{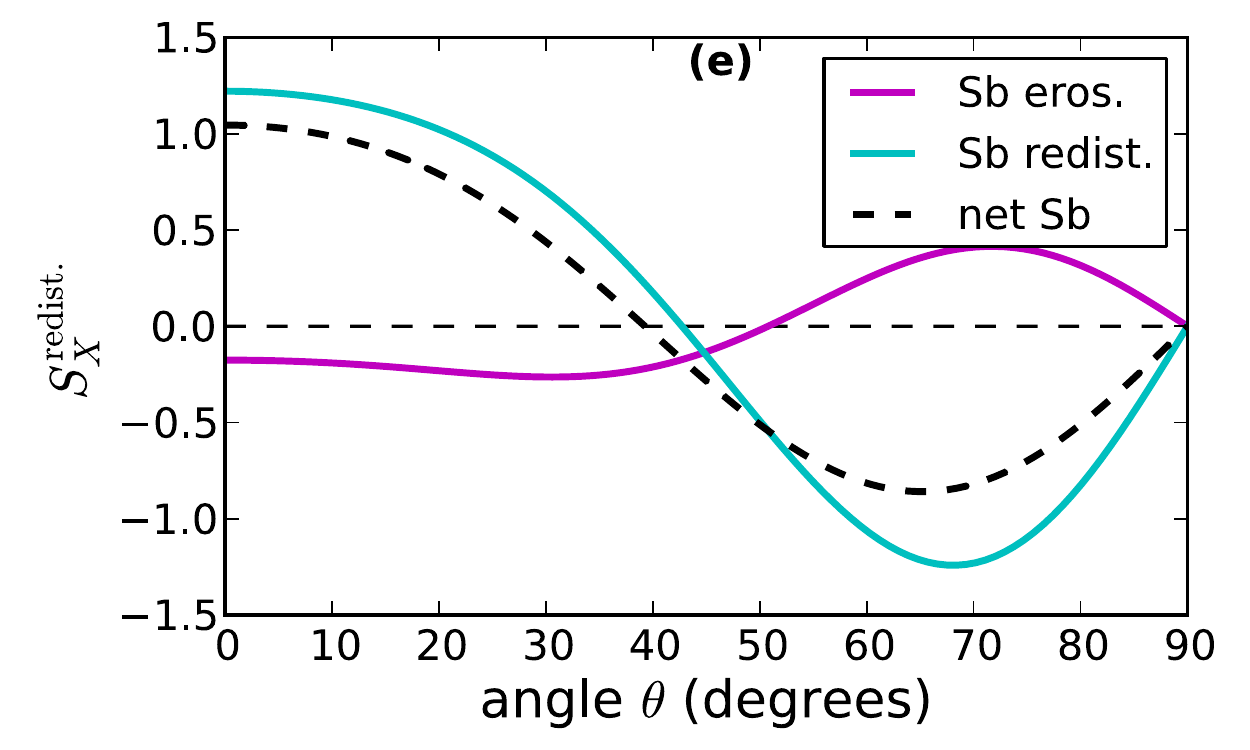}\includegraphics[width=2.15in]{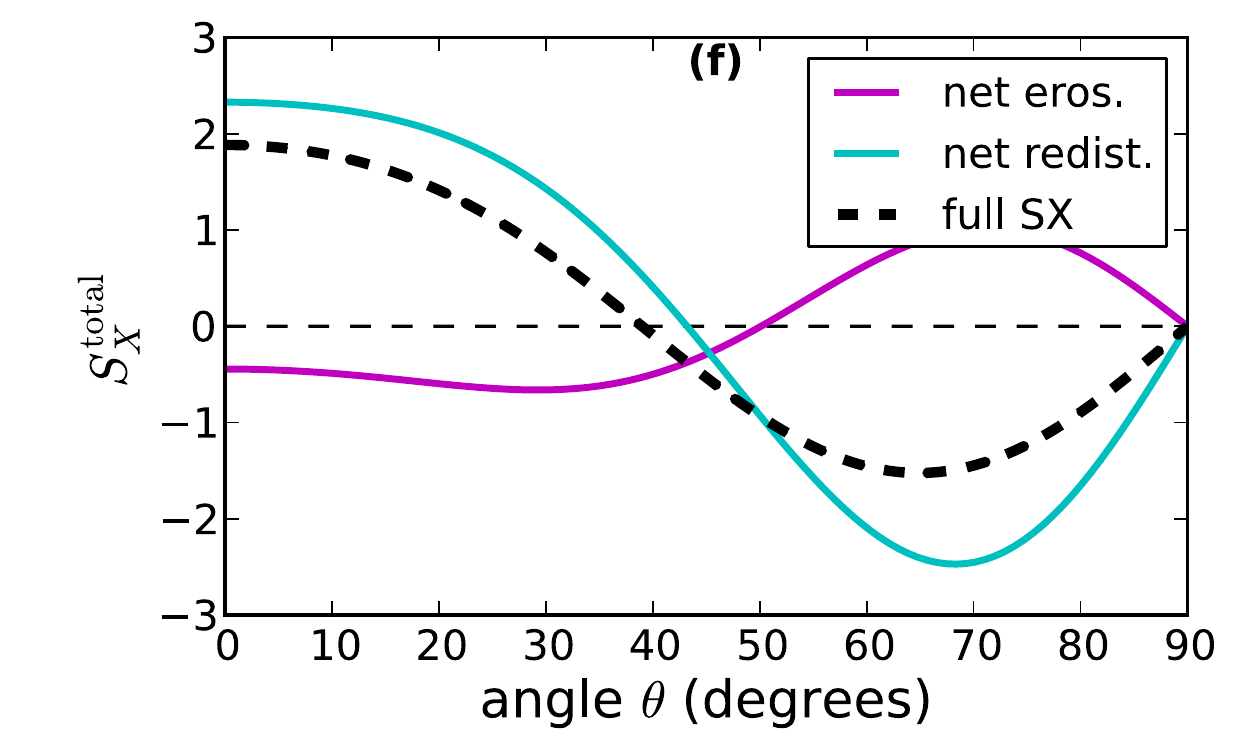}
\par\end{centering}

\begin{centering}
\includegraphics[width=2.15in]{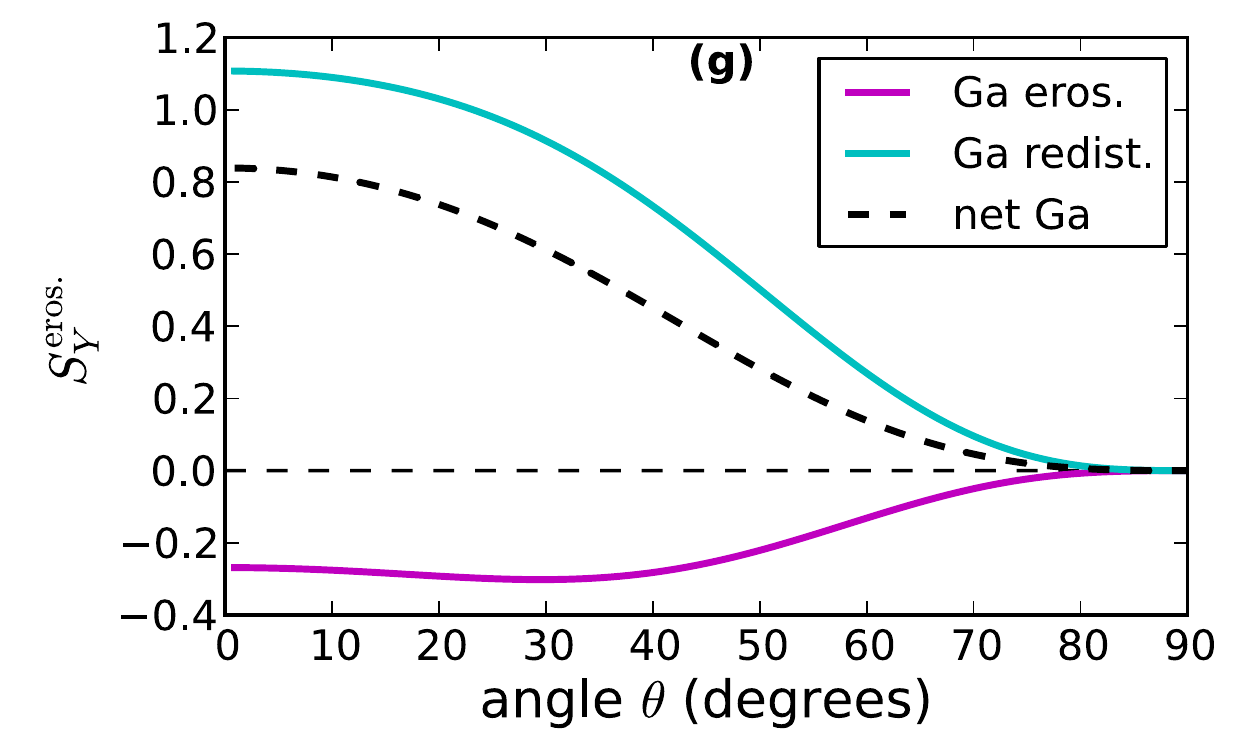}\includegraphics[width=2.15in]{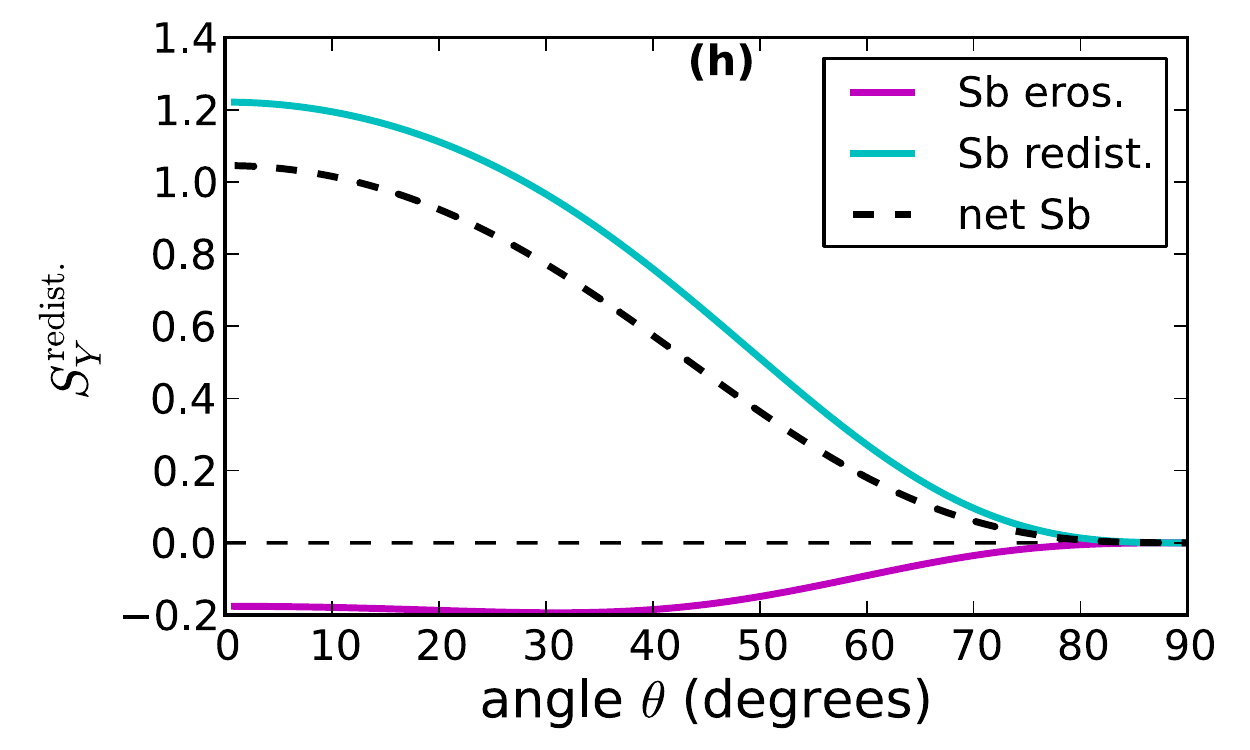}\includegraphics[width=2.15in]{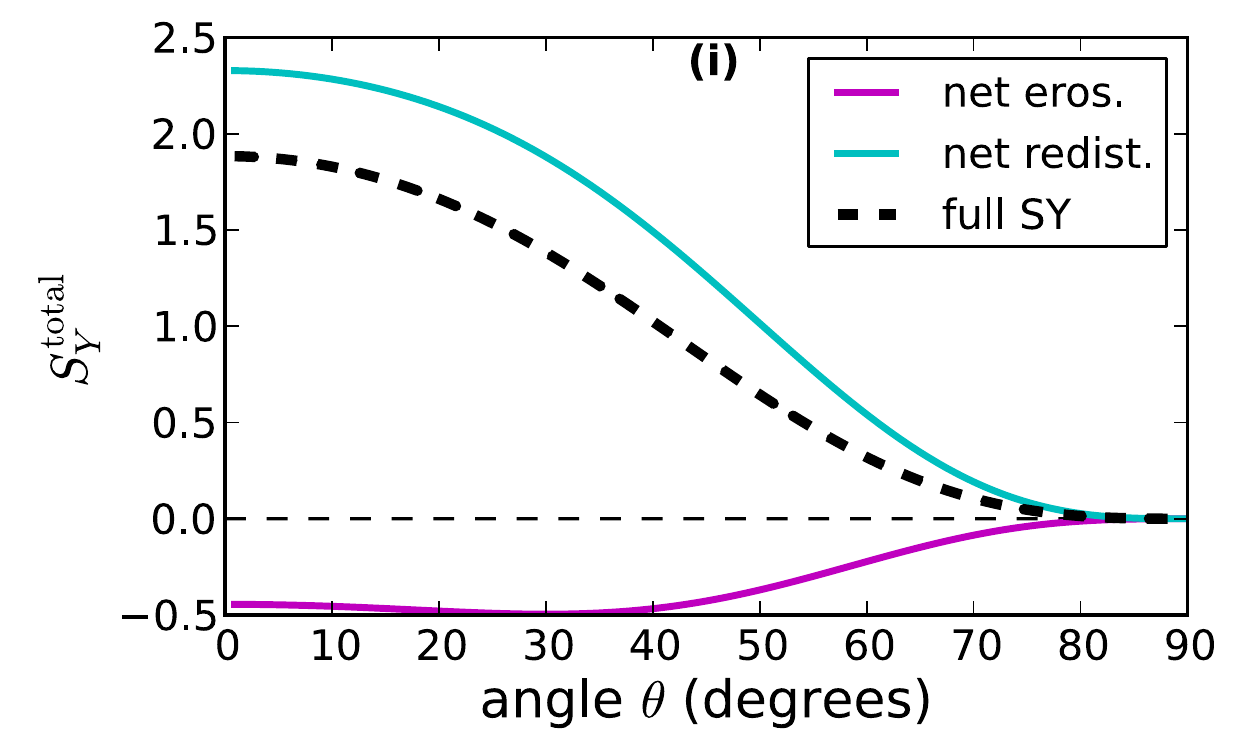}
\par\end{centering}

\begin{centering}

\par\end{centering}

\centering{}\caption{Moments and coefficients by component. (a-c) The zeroth moment $M^{\left(0\right)}$,
the first erosive moment $M_{\mathrm{eros.}}^{\left(1\right)}$, and
the first redistributive moment $M_{\mathrm{redist.}}^{\left(1\right)}$;
decomposed into contributions from Ga and Sb. These plots indicate
that Ga is preferentially sputtered from a homogeneous, amorphous
film at 50/50 initial concentrations, but that both species are redistributed
approximately equally. (d-f) The erosive and redistributive contribution
to $S_{X}$ for each species separately, and summed over both species.
(g-i) The corresponding plots for $S_{Y}$ (for completeness only
- at $\theta=0$ both values are equal). It is readily seen that the
contributions of erosion and redistribution have the same relative
shapes and sizes for each species, with the redistributive component
playing the dominant role at most angles. Note that while the full
angle-dependence of the curves in (d-i) is estimated for illustrative
purposes, our concern here is only with the values at $\theta=0$.
\label{fig: moments}}
\end{figure}

\par\end{center}

The results of our simulations are shown in Fig.\ref{fig: moments}.
We first calculate the moments according to Eq.(\ref{eqn: moments-from-positions}),
at nine different angles, and plot them along with three-term Fourier
fits. For comparison with the normal-incidence irradiation conditions
of the Bradley-Shipman theory, our primary interest is in the simulations
near $\theta=0$. However, additional data at intermediate angles
is also provided for general reference, and these, along with symmetry
considerations at $\theta=90$, improve the quality of the fits (see
Ref.\citep{norris-etal-NCOMM-2011} for more details).\textcolor{black}{{}
Our Fig.~\ref{fig: moments}a indicates that, for an initial fresh,
unsegregated 50/50 target, Ga is preferentially sputtered relative
to Sb, and consequently exhibits a stronger curvature dependent sputtering
effect in Fig.~\ref{fig: moments}b (see Section~\ref{sub:Limitations}
below for a discussion of this result). In contrast, the strikingly
similar shape of the curves in Fig.~\ref{fig: moments}c suggests
that preferential redistribution is unlikely to be a significant effect
for this system. A comparison of the relative magnitudes of the curves
in Figures~~\ref{fig: moments}b and ~\ref{fig: moments}c suggests
that redistribution will overwhelm the effect of erosion, as has been
seen in pure materials. }

\textcolor{black}{To confirm that these impressions are not altered
under the derivatives in Eqs.(\ref{eqn: moments-to-coeffs}), and
to obtain exactly the coefficients required by Eq.(\ref{eqn: craters-to-BS-theory})
for comparison with the BS theory, we also plot the coefficients found
in that equation in Figures~\ref{fig: moments}d-i. Here we compare
the erosive and redistributive contributions to both $S_{X}$ and
$S_{Y}$ for each atomic species separately, and the net erosive and
redistributive contributions summed over both species. We confirm
that the redistributive contributions are dominant, and we also note
the striking similarity of the shapes of the curve for each species,
which are also similar to the shapes observed in pure silicon \citep{norris-etal-NCOMM-2011}.
This indicates that chemical composition is not a strong influence
on the collision cascade in a homogeneous target, suggesting that
the dominance of redistribution may potentially be a generic feature
of low-energy ion bombardment.}

Using the data shown in Figure~\ref{fig: moments}, and the definitions
of the coefficients from Eqs.(\ref{eqn: moments-to-coeffs}), we now
proceed to obtain the values of the coefficients of Equation~(\ref{eqn: pure-h-equation})

\begin{equation}
\begin{aligned}Y^{\mathrm{Ga}} & = & -0.0172\times I_{0}\,\frac{\mathrm{nm}^{\phantom{2}}}{\mathrm{s}}\\
Y^{\mathrm{Sb}} & = & -0.0102\times I_{0}\,\frac{\mathrm{nm}^{\phantom{2}}}{\mathrm{s}}\\
S_{X,Y}^{\mathrm{Ga,eros.}} & = & -0.269\times I_{0}\,\frac{\mathrm{nm}^{2}}{\mathrm{s}}\\
S_{X,Y}^{\mathrm{Sb,eros.}} & = & -0.176\times I_{0}\,\frac{\mathrm{nm}^{2}}{\mathrm{s}}\\
S_{X,Y}^{\mathrm{Ga,redist.}} & = & 1.11\times I_{0}\,\frac{\mathrm{nm}^{2}}{\mathrm{s}}\\
S_{X,Y}^{\mathrm{Sb,redist.}} & = & 1.22\times I_{0}\,\frac{\mathrm{nm}^{2}}{\mathrm{s}}
\end{aligned}
\label{eqn: craters-to-BS-theory-1}
\end{equation}
where all coefficients are estimated at $\theta=0$ and $c_{A}=c_{B}=0.5$,
and where $I_{0}$ is again the flux in the beam direction, as in
Eq.(\ref{eqn: moments-to-coeffs}).

\subsection{Extrapolation to Steady Concentration}

The final step in estimating the coefficients $\left\{ A,C,A^{\prime},C^{\prime}\right\} $
is to observe that the various components of the moments $M^{\left(0\right)}$
and $M^{\left(1\right)}$ are functions not only of angle, but also
of the concentrations of the various species within the target. Similarly,
within the BS theory, the quantity $\Lambda_{i}$ - associated with
sputter yield - and presumably also $\mu_{i}$ - describing the redistributive
flux - share this dependence. {[}The latter dependence was not discussed
in Ref.\citep{shipman-bradley-PRB-2011}, but presumably exists.{]}
In both Equations (\ref{eqn: craters-to-BS-theory}) and (\ref{eq: craters-to-BS-conversion}),
therefore, all coefficients $Y$ and $S$ must in principle be evaluated
at the steady-state concentration $\left(c_{A,0},c_{B,0}\right)$.
The steady concentration, in turn, depends implicitly upon the relative
shapes of the yield functions $\Lambda_{A}\left(c_{A}\right)$ and
$\Lambda_{B}\left(c_{B}\right)$, as specified in Eq.~(12) of Ref.\citep{shipman-bradley-PRB-2011}.
This presents a challenge to a molecular dynamics investigation, because
in addition to the simulations over many angles described here, one
would also have to perform simulations over many different target
compositions. This, in turn, requires constructing many different
targets, which is actually the most time-consuming step. 

Because of these considerations, and limited computational capacity,
we are led to make an important simplifying assumption. We will assume
that each species both sputters and redistributes at a rate linearly
proportional to its concentration in the film (c.f. Refs. \citep{bradley-PRB-2011c,bradley-JAP-2012}),
with rate zero when its concentration is zero, and rate equal to our
measured rate when its concentration is one half. This gives, for
the moments,
\begin{equation}
\begin{aligned}M_{Z}^{\left(0\right)}\left(c_{Z}\right) & =M_{Z}^{\left(0\right)}\left(\frac{1}{2}\right)\times\left(2c_{Z}\right)\\
M_{Z,\mathrm{type}}^{\left(1\right)}\left(c_{Z}\right) & =M_{Z,\mathrm{type}}^{\left(1\right)}\left(\frac{1}{2}\right)\times\left(2c_{Z}\right)
\end{aligned}
\label{eqn: moment-assumptions}
\end{equation}
and for the corresponding assumption on the coefficients,

\begin{equation}
\begin{aligned}Y^{Z}\left(c_{Z}\right) & =Y^{Z}\left(\frac{1}{2}\right)\times2c_{Z}\\
S^{Z,\mathrm{type}}\left(c_{Z}\right) & =S^{Z,\mathrm{type}}\left(\frac{1}{2}\right)\times2c_{Z}
\end{aligned}
.\label{eqn: Lambda-mu-estimate}
\end{equation}
Under this assumption, we will be able both to estimate the steady
concentration, and also estimate coefficients at that concentration,
without excessive numerical simulation (see Section~\ref{sub:Limitations}
below for further discussion).

Under the assumptions (\ref{eqn: moment-assumptions}) and (\ref{eqn: Lambda-mu-estimate}),
and using Eq.~(12) of Ref.\citep{shipman-bradley-PRB-2011}, the
values in Eq.(\ref{eqn: craters-to-BS-theory}) for $Y^{\mathrm{Ga}}$
and $Y^{\mathrm{Sb}}$ predict steady concentrations of Ga and Sb
to be 
\[
\begin{aligned}c_{\mathrm{Ga},0} & =0.37\\
c_{\mathrm{Sb},0} & =0.63
\end{aligned}
\]
Then, combining the estimates (\ref{eqn: craters-to-BS-theory-1})
at the 50/50 concentration with the assumption (\ref{eqn: Lambda-mu-estimate})
on the concentration dependence of the coefficients, we use expression
(\ref{eq: craters-to-BS-conversion}) to produce final estimates for
$\left\{ A,C,A^{\prime},C^{\prime}\right\} $ of

\begin{equation}
\begin{aligned}A & \approx & 0.014\times I_{0}\,\frac{\mathrm{nm^{\phantom{2}}}}{\mathrm{s}}\\
C & \approx & 1.9\times I_{0}\,\frac{\mathrm{nm}^{2}}{\mathrm{s}}\\
A^{\prime} & \approx & .0091\times I_{0}\,\frac{\phantom{2}1\phantom{2}}{\mathrm{s}}\\
C^{\prime} & \approx & -.12\times I_{0}\,\frac{\mathrm{nm}^{\phantom{2}}}{\mathrm{s}}
\end{aligned}
.\label{eq: parameter-value-results}
\end{equation}
where we have estimated the value $\Delta\approx3\,\mathrm{nm}$ from
TEM measurements of pure silicon irradiated at the same 250 eV \citep{madi-thesis-2011}.

\subsection{Discussion}

A major accomplishment of the Bradley-Shipman theory was its identification
of a narrow band of unstable wavenumbers in the Equations (\ref{eqn: linear-height-evolution-BH})-(\ref{eqn: linear-concentration-evolution-BH}),
bounded both above and below by stable wavenumbers. As described in
Section~\ref{sub: finite-wavelength-requirements}, this is the essential
first ingredient for producing highly-ordered patterns of any kind,
and the BS theory was the first physically-derived theory to admit
this kind of instability. However, to do so it placed quite severe
requirements on the parameters in the model. 

We quickly summarize these constraints by recalling the generic dispersion
relation 
\begin{equation}
\sigma_{+}\left(k\right)=\frac{1}{2}\left(-\tau+\sqrt{\tau^{2}-4\Delta}\right).\label{eqn: quadratic-solution-of-sigma-1}
\end{equation}
and the coefficients defined in the BS theory

\begin{eqnarray}
\tau\left(k\right) & = & A^{\prime}+\left(C+B^{\prime}\right)k^{2}+Dk^{4}\label{eqn: tau-1}\\
\Delta\left(k\right) & = & \left(CA^{\prime}-AC^{\prime}\right)k^{2}+\left(CB^{\prime}+DA^{\prime}\right)k^{4}+\left(DB^{\prime}\right)k^{6}.\label{eqn: Delta-1}
\end{eqnarray}
Now, a general feature of Eq.(\ref{eqn: quadratic-solution-of-sigma-1})
is that a mode is only unstable if either $\tau<0$ or $\Delta<0$
\citep{cross-greenside-2009-book}. As described above, a narrow band
of unstable modes requires stable longwaves, which in turn requires
that $\left(CA^{\prime}-AC^{\prime}\right)>0$. Then, because the
parameters $D$, $A^{\prime}$, and $B^{\prime}$ are all positive,
we see that the only way to achieve $\tau<0$ or $\sigma<0$ is if
(a) the coefficient $C$ is sufficiently negative to drive the instability,
which in turn requires that (b) the parameter group $AC^{\prime}$
is sufficiently negative to stabilize the long waves, with $\left|AC^{\prime}\right|>\left|A^{\prime}C\right|$.
These were striking requirements of the Bradley-Shipman instability,
as the former requirement, in particular, is contrary to results for
pure Silicon and Germanium \citep{madi-etal-PRL-2011,anzenberg-etal-PRB-2012},
where $C>0$ due to the dominance of redistributive over erosive effects
\citep{norris-etal-NCOMM-2011}. 

Indeed, as we have shown here, (1) the positive value of $C$ persists
in the case of binary GaSb, in a way that suggests this may be a generic
feature of low-energy ion irradiation; (2) the parameter group $AC^{\prime}$
in Eq.(\ref{eqn: Delta}) (the only place that either $A$ or $C^{\prime}$
occur in the BS theory) is much smaller in magnitude than the similarly-dimensioned
group $A^{\prime}C$, suggesting that preferential redistribution
is likely not an important effect. These findings are contrary to
the requirements of the Bradley-Shipman theory - in fact, because
both $C$ and $\left(A^{\prime}C-C^{\prime}A\right)$ were found to
be positive, and because $D$, $A^{\prime}$, and $B^{\prime}$ are
positive by definition, the BS theory should predict stable, flat
surfaces for normal-incidence irradiation of GaSb, in contrast to
the dot-like patterns seen experimentally.

\subsection{Limitations.\label{sub:Limitations}}

We want to acknowledge the following limitations of the MD simulations
performed in Section~\ref{sub: environment-and-results}, due to
finite computational resources in the simulation stage. Although they
are not an inherent limitation of the general approach described in
Section~\ref{sec: Coefficient-Estimation}, they provide directions
for future improvement of our estimates.

First, the definitions for the coefficients $Y^{Z}$ and $S^{Z,\mathrm{type}}$
in Eq.(\ref{eqn: moments-to-coeffs}), like the corresponding definitions
in Ref.\citep{norris-etal-NCOMM-2011} from which they are generalized,
are the result of performing simulated impacts on an initially flat
target surface. As such, they represent simplifications of the more
general results in Ref.\citep{norris-etal-2009-JPCM} for arbitrary
curved surfaces. Obtaining such generalized results would, unfortunately,
require simulation of impacts not only over many angles, but also
on many different types of curved surface, which is outside our current
computational capacity. Although this simplification may change the
absolute magnitude of the identified coefficients, it does not significantly
change the relative magnitudes, upon which our conclusions are entirely
based.

Second, our simulations were performed only at the 50/50 concentration
for a fresh, homogeneous target. We have estimated values of coefficients
at the steady concentration by means of the linear approximations
in Eqs. (\ref{eqn: moment-assumptions})-(\ref{eqn: Lambda-mu-estimate}),
as exemplified in a similar context \citep{bradley-PRB-2011c}. Like
the flat-surface simplification, this avoids the simulation of targets
at a variety of concentrations. With more computational resources,
such simulations could of course be performed, which would improve
upon the accuracy of the estimates. However, in addition to the computational
cost of the simulations themselves, this would require the construction
of many targets, which is in fact the most time-consuming part of
our study. In addition, the nature of the irradiated target away from
the 50/50 concentration is not currently well-known. 

Third, our results indicating that Ga is sputtered preferentially
relative to Sb for a fresh, unsegregated target at the 50/50 concentration
do not agree with early observations of significant Ga \emph{enrichment}
at the surface \citep{le-roy-etal-JAP-2009}, even (to a lesser extent)
in the absence of oxygen which preferentially oxidizes Ga at the surface
\citep{el-atwani-etal-JAP-2011,el-atwani-etal-APL-2012a,el-atwani-etal-NIMB-2012}.
Because of the disagreement with experiment, we also performed simulations
in SDTrimSP \citet{mutzke-etal-2011-sdtrimsp-5}. For a variety of
inter-atomic potentials, these simulations were roughly consistent
with the results from MD, with around a 10\% preferential yield of
Ga at the 50/50 composition, and a significantly higher yield of pure
Ga than for pure Sb. These atomistic results are all consistent with
the more efficient energy transfer of Ar to Ga relative to the heavier
Sb. It is therefore likely that experimental target conditions differ
from the ones in typical atomistic targets in ways that still need
to be carefully understood. For instance, although the methods used
to produce amorphous GaSb discussed above result in an approximately
homogeneous target, in Refs.\citep{yu-sullivan-saied-SS-1996-gibbsian-segregation,el-atwani-etal-NIMB-2012}
it was found that the lower vacuum surface energy of Sb may pull these
atoms preferentially to the top monolayer of the target via Gibbsian
segregation, resulting in a surface layer enriched with Sb above a
subsurface layer enriched with Ga. It is easy to speculate that such
a configuration may lead to enhanced sputtering of Sb relative to
an unsegregated target, and preliminary SDTrimSP simulations suggest
that this could readily promote Sb to be the preferentially sputtered
species. In any case, it is clear that future work to bring the simulated
results into agreement with experiment will require a better understanding
of precisely what is happening during the early stages of irradiation,
and the construction of an atomistic target reflecting that understanding.

Despite the limitations just described, we anticipate that our two
principal conclusions for the GaSb system are likely to remain valid
as more accurate targets are developed. In particular, although the
relative response of Ga vs. Sb atoms might well be altered by surface
segregation, it is more doubtful that the relative response of eroded
vs redistributed atoms would be so altered -- Figure~\ref{fig: moments}
demonstrates that redistribution overwhelms erosion for each species
\emph{individually}, precisely as observed in pure Silicon at similar
energies \citep{norris-etal-NCOMM-2011}. Therefore, the contributions
to the sign of $C$ are positive \emph{for both species}, and changing
the relative amount of Ga and Sb in the target would not change the
sign of this parameter. Regarding the relative importance of preferential
redistribution, the parameters $A$ and $C^{\prime}$ - which depend
on \emph{differential} erosion and redistribution - may be expected
to vary somewhat for different target structures. However, these only
appear as the product $AC^{\prime}$, which only appears together
with the group $A^{\prime}C$. The latter group depends only on \emph{net}
erosion and redistribution, and should therefore be altered much less
by compositional changes. Because this latter group was found to be
ten times larger than the former, we suspect that even improved estimates
with better targets would be unlikely to reverse the relative sizes
of these terms.

\section{Conclusions}

We have shown that many parameters of coupled PDE models for the irradiation
of two-component materials may be estimated using Molecular Dynamics
simulation by means of an extension, to binary materials, of the theory
of crater functions described in \citep{norris-etal-2009-JPCM,norris-etal-NCOMM-2011}.
In principle, these include parameters describing net sputtering and
redistribution ($C$), preferential sputtering ($A$), preferential
redistribution ($C^{\prime}$), and material replenishment ($A^{\prime}$).
Remarkably, these are exactly the four parameters needed to calculate
the critical parameter group $\left(A^{\prime}C-C^{\prime}A\right)$,
which governs the behavior of long waves in all such theories. The
methods demonstrated here are general, and should therefore be applicable
to a wide variety of coupled PDE models for irradiated binary systems
\citep{bradley-shipman-PRL-2010,shipman-bradley-PRB-2011}, enabling
the prediction of this important distinguishing characteristic among
a variety of pattern-forming systems.

Subsequently, we have performed the first estimate of these four parameters
within the Bradley-Shipman theory of irradiated binary alloys, as
applied to the low-energy irradiation of GaSb by Ar, in which hexagonal
dot arrays were first observed. Our main finding for this system is
that the parameter $C$ is positive for normal incidence irradiation
of GaSb, indicating the dominance of redistribution over erosion,
as was predicted by MD for pure Si \citep{norris-etal-NCOMM-2011},
and shown experimentally for both pure Si and Ge \citep{madi-etal-PRL-2011,anzenberg-etal-PRB-2012}.
Additionally, we saw that the parameter $C^{\prime}$ associated with
preferential redistribution, hypothesized to be an important physical
effect in ordered dot formation, appears too small to play an important
role. 

Both the positive value of $C$, and the relative unimportance of
$C^{\prime}$, are contrary to requirements of the Bradley-Shipman
theory on the formation of well-ordered patterns. In fact, for the
parameter values we have estimated, that theory would predict entirely
smooth surfaces for low-energy, normal-incidence irradiation of GaSb.
Especially if our results extend to a wide range of other two-component
materials (in other words, if they are a generic property of the kinetically-dominated
collision cascade for low energy irradiation), then these findings
provide strong motivation for the consideration of alternative physical
mechanisms to explain the ordered patterns observed in these systems.
We investigate such a mechanism elsewhere \citep{norris-arXiv-2013-linear-phase-separation}.

\bibliographystyle{elsarticle-num}
\bibliography{/home/snorris/Dropbox/research/bibliography/tagged-bibliography,/home/scott/Dropbox/research/bibliography/tagged-bibliography}

\end{document}